# Brain Computer Interface Technology for Future Battlefield

Guodong Xiong, Xinyan Ma, Wei Li, Jiaqi Cao, Jian Zhong, Yicong Su

China South Industries Group Corporation, Hangzhou Zhiyuan Research Institute Co., Ltd., 310012

Email: jiaqicao.pd@gmail.com

*Abstract*—With the development of artificial intelligence and unmanned equipment, human-machine hybrid formations will be the main focus in future combat formations. With the development of big data and various situational awareness technologies, while enhancing the breadth and depth of information, decision-making has also become more complex. The operation mode of existing unmanned equipment often requires complex manual input, which is not conducive to the battlefield environment. How to reduce the cognitive load of information exchange between soldiers and various unmanned equipment is an important issue in future intelligent warfare. This paper proposes a brain computer interface communication system for soldier combat, which takes into account the characteristics of soldier combat scenarios in design. The stimulation paradigm is combined with helmets, portable computers, and firearms, and brain computer interface technology is used to achieve fast, barrier free, and hands-free communication between humans and machines. Intelligent algorithms are combined to assist decision-making in fully perceiving and fusing situational information on the battlefield, and a large amount of data is processed quickly, understanding and integrating a large amount of data from human and machine networks, achieving real-time perception of battlefield information, making intelligent decisions, and achieving the effect of direct control of drone swarms and other equipment by the human brain to assist in soldier scenarios.

*Keywords—Brain computer interface, Intelligent warfare, Human-computer interaction*

## I. Introduction

In the near future, with the development of technology, the essence of war will undergo significant changes. The battlefield is no longer a traditional battle between soldiers but has become a technological battle in which human soldiers and unmanned equipment work together. In most cases, soldiers are no longer at the forefront of combat but have become highly skilled operators. From aerial vehicles to ground robots, all kinds of unmanned equipment have brought about the progressiveness and diversity of operations, as well as the complexity of operations. Existing traditional communication systems usually require manual input, such as pressing a button or using voice commands [2], which has the disadvantages of large information transmission delay, low security and anti-interference performance. In the high-pressure environment of soldier operations, the user interface and human-machine interaction of equipment play a crucial role in communication. The user interface of the device should be intuitive and easy to use, minimizing the cognitive load between soldiers and devices.

Brain computer interface technology can achieve hands-free information exchange. The signal input method is not traditional button input or voice input, but directly collects the EEG signals generated by the human brain and converts them into mechanically recognizable instructions. It can eliminate the cognitive load between soldiers and unmanned equipment operations, accelerate the operator's "observe, orient, decide, and act" (OODA) cycle. The existing brain computer interface technology is mainly used in the medical field [5], such as implanting electrodes in paralyzed patients with spinal cord injury [6], using brain computer interface technology to control robotic arms to complete specific actions [7], or strengthening brain responses through neural feedback training [8], analysing participant EEG signals and stimulating them with visual and auditory signals for positive feedback. There is currently a lack of mature products for brain computer interface technology in soldier combat communication, and related applications still need to be developed.

The existing brain computer interface technology has the problem of insufficient accuracy. Although non-invasive brain computer interface technology has the advantages of being easy to wear and use without the need for electrodes, the collected signals contain a large amount of noise [9]. Therefore, the control accuracy and precision applied to communication between soldiers and external unmanned equipment are insufficient, and the integration between brain computer interface technology and existing communication technology is not sufficient, lacking standardized protocols [10]. The above reasons have led to the fact that brain computer interface technology has not yet reached the level of industrialization, especially for high-performance and low latency brain computer interface communication between soldiers and external unmanned devices.

This paper will propose a brain computer interface system for communication between soldiers and unmanned intelligent agents, achieving efficient communication between operators and unmanned equipment. This system can provide integration between soldiers and unmanned equipment, promote seamless information exchange between users and unmanned equipment groups, and reduce cognitive load with unmanned equipment. This integration will improve the effectiveness and versatility of unmanned equipment, enhance user control over unmanned equipment, enable them to collaborate on more complex tasks, and thus achieve a high degree of human-machine collaboration.

## II. Method

### A. EEG signal acquisition

Considering the difficulty of combining the collection device with the soldier's helmet, non-invasive collection of soldier's EEG signals (EEG) is adopted as the signal input, and the EEG electrodes use a more comfortable flat electrode. This approach has advantages such as low cost, low risk, simple operation, and ease of use [11], and commonly used EEG collection devices such as electrode caps fit well with the head, making it easy to integrate with the soldier's helmet.

*B. Stimulation paradigm*

Use a stimulation paradigm based on steady-state visual evoked potential (SSVEP) in the stimulation method and implement frequency encoding. The SSVEP paradigm refers to the potential changes induced in the visual cortex of the brain that are consistent with the stimulus frequency and its harmonic frequency when the user focuses on a constant frequency of flashing stimuli [12]. This paradigm has a clear periodicity consistent with the stimulus frequency, strong recognizability, and a higher signal-to-noise ratio than other evoked potentials. Using stimuli of different frequencies can generate several different types of EEG signals to correspond to control commands from various external devices. The minimum frequency required by the SSVEP paradigm is 6Hz, and low-frequency stimuli below 15Hz elicit the strongest response [13]. Therefore, the optimal SSVEP stimulation frequency should be selected between 6Hz and 15Hz. The SSVEP paradigm can be well integrated with smart equipment carried by soldiers, such as displaying stimulus paradigms through screens located on helmets or portable smart devices, or installing stimulus paradigms on firearms. Soldiers can induce EEG signals of different frequencies by staring at stimuli at different positions and frequencies, corresponding to different mechanically recognizable instructions.。

*C. Lead electrode*

In terms of lead selection, due to the fact that the EEG response caused by visual evoked potentials is mainly distributed in the occipital lobe area [14], the occipital lobe electrode is used as the acquisition channel for SSVEP EEG signals, the CPz in the parietal lobe is used as the reference electrode, and the AFz in the prefrontal lobe is used as the grounding electrode. Using a head mounted EEG conductive electrode cap as an EEG acquisition device, it fits well with the scalp and is convenient for integrated design with soldier helmets.

*D. Signal pre-processing*

After completing the collection of EEG signals, due to the presence of unnecessary frequency components and noise in the EEG signals, there is interference in the feature extraction and recognition of the signals. Therefore, preprocessing of the collected EEG signals is necessary, including signal filtering, denoising, and amplification. The pre-processed signal can be used for feature extraction and pattern recognition, corresponding different stimulus paradigms to different mechanically recognizable instructions. Due to the majority of lead selection using the occipital lobe electrode, which preserves most of the EEG responses caused by visual evoked potentials, spatial filtering has been achieved to some extent, removing eye movement artifacts with higher responses on the frontal lobe electrode [15].。

*E. Signal feature extraction and recognition*

In terms of signal feature extraction and recognition, the typical correlation analysis method (FBCCA) combined with filter banks is adopted [16]. The filter bank (FB) uses multiple bandpass filters to divide the collected signal into multiple sub frequency bands in the frequency domain for subsequent canonical correlation analysis (CCA). CCA is a multivariate statistical method for studying the correlation between variables, commonly used for feature recognition in the SSVEP paradigm [17]. This method can identify the correlation between multi-channel EEG signals extracted by the SSVEP paradigm and sine and cosine signals, transforming multivariate correlation analysis into univariate correlation analysis. FB is a method of decomposing a signal into signals on different sub bands using multiple bandpass filters with different cutoff frequencies. It is commonly used to analyse signals containing multiple frequency band components. Due to the fact that the harmonic components of SSVEP can provide good robustness information to the input EEG signal, the high-frequency cutoff frequency in the filter bank can be set as the fourth harmonic component of the highest stimulation frequency of SSVEP. The low-frequency cutoff frequencies are designed as the starting frequencies of each harmonic component.

## III. SYSTEM DESIGN

This chapter will introduce the design of the proposed brain computer interface system, which enables efficient communication between operators and unmanned equipment. Human brain consciousness (EEG signals) is used to control external unmanned equipment, and feedback signals are obtained from external equipment to transmit to users for real-time perception of on-site information and intelligent decision-making. The human brain directly controls unmanned equipment such as drone swarms to achieve auxiliary combat effects.

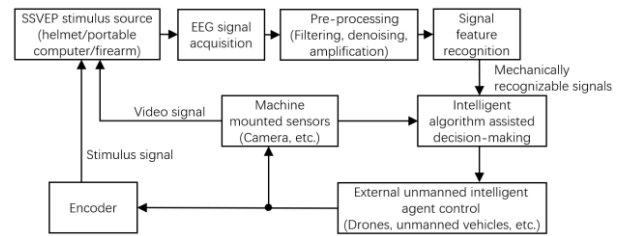

Figure 1. The structural design of the proposed brain computer interface system

Install the display area that presents the SSVEP paradigm in a location that is convenient for soldiers to observe during combat, such as helmet displays, portable individual computers, or deploying LED modules on firearms. The SSVEP paradigm with different flicker frequencies will induce corresponding frequency EEG signals in the soldier's visual cortex to correspond to different mechanically recognizable instructions. After the collection of EEG signals, it is necessary to preprocess the collected signals, including filtering, denoising, and amplification. Filtering can remove unnecessary frequency components from EEG signals, and denoising and amplification can enhance the recognizability of collected EEG signals. The pre-processed EEG signals will be used for feature recognition using the FBCCA method. First, a filter bank will be used to group the entire EEG signal according to the cutoff frequency. Then, the typical correlation analysis method will be used to map the grouped EEG signals one by one to the discretized results.

In addition, there are fundamental limitations in using different paradigms to correspond to different mechanical instructions. A single instruction can only correspond to a simple control of one type of external device, such as the simple movement of a drone in space, forward and backward left and right movements, or a set of instructions for a series of actions. Faced with the complex environment on the battlefield, it is difficult to achieve the tactical needs of future battlefields through simple behaviour control of external devices achieved through command mapping. Moreover, the combat situation on future battlefields is highly complex and constantly changing, making it difficult for the human brain to quickly recognize and make decisions. Therefore, in this design, an intelligent algorithm assisted decision-making module is introduced, which is installed in various unmanned equipment and uses methods such as image recognition algorithms Basic algorithms such as reinforcement learning, combined with databases such as military targets, can achieve the effect of assisting soldiers in combat. The EEG signals emitted by soldiers are pre-processed and feature recognized and converted into mechanically recognizable instructions as input into the auxiliary decision-making module. This module combines input information from external devices such as onboard cameras and radars to intelligently perceive and fuse massive information on the battlefield, mining key information such as enemy position, enemy type, and surrounding terrain, And provide intelligent decision-making based on the control instructions given by the soldiers. If a soldier gives a "regional reconnaissance" command to a reconnaissance drone through EEG signals, the drone will call a map construction algorithm to scan and construct a map of the nearby terrain. Combined with intelligent path planning algorithms, the drone will patrol the surrounding terrain and use facial recognition or military target recognition algorithms to recognize and record enemy infantry or tanks and other targets. The intelligent algorithm decision-making module based on mechanical learning can also be continuously optimized and upgraded through the accumulation of massive data and experience, improving the correctness and accuracy of operations on the battlefield. Soldiers in complex and ever-changing battlefields only need to provide "fuzzy" control instructions to external devices through brain computer interface technology to achieve precise automation control of unmanned equipment groups.

The battlefield situation information captured by sensors of external unmanned equipment will be transmitted to combat soldiers through feedback loops, helping them analyse the battlefield situation and making further decisions. The feedback loop can provide two types of battlefield situation feedback information. One is to transmit the video stream captured by the cameras on the unmanned equipment group being controlled to the intelligent devices with displays worn by soldiers, helping them obtain battlefield information and assisting them in combat. The second is to use an encoder to reverse encode the battlefield situation information, as well as the recognition and execution status of the previous command, in the form of a stimulus signal reflected in the display system of the soldier's helmet. This stimulus signal is similar to the SSVEP stimulus paradigm during EEG collection, and the recognition status of the previous EEG signal is considered as unsuccessful recognition, Then, the stimulus signal presented on the helmet display is returned as a red flashing block through the encoder; If the recognition of the previous EEG signal was successful but not executed successfully, such as when commanding a drone to perform a strike task and successfully recognizing the EEG signal but not successfully striking, the stimulus signal displayed on the helmet display is returned as a yellow flashing block through the encoder; If the recognition of the previous EEG signal is successful and executed successfully, the encoder returns a green flashing block of the stimulus signal displayed on the helmet display.

IV. CONCLUSION

The existing brain computer interface technology is mostly used in the field of medical rehabilitation, and there is still room for development in military applications. And the existing traditional military communication systems usually require manual input, such as pressing buttons or using voice commands, which has disadvantages such as large information transmission delay, low security and anti-interference performance. This article proposes a brain computer interface system for soldier combat scenarios, which can achieve efficient communication between soldiers and external unmanned intelligent agents. In terms of design, non-invasive brain electrode caps are used to collect soldier's EEG signals, facilitating integration with soldier helmets. The SSVEP stimulation paradigm is used in terms of stimulation methods, which can be deployed on helmets, portable computer displays, or firearms for easy observation and use during combat. After preprocessing and feature recognition, EEG signals are mapped to mechanically recognizable instructions, and battlefield situation information perception and feedback are achieved through intelligent algorithms deployed by external unmanned devices. Unlike traditional communication technologies, this brain computer interface system can achieve fast, barrier free, and hands-free communication between humans and machines, and fully perceive and integrate situational information on the battlefield. The rapid processing of a large amount of data, understanding and synthesizing a large amount of data from human and machine networks, can achieve real-time perception of battlefield information, make intelligent decisions, and achieve direct control of unmanned intelligent agents by the human brain, achieving high cooperation between humans and machines, Enable soldiers and unmanned equipment to collaborate on more complex tasks.